\documentclass[a4paper,aps,pra,preprintnumbers]{revtex4}
\usepackage[utf8]{inputenc}
\usepackage[english,russian]{babel}
\usepackage[T1,T2A]{fontenc}
\usepackage{amssymb,amsfonts,amsmath,mathtext,enumerate,float,dsfont}
\usepackage{graphics,graphicx,epsfig,epstopdf}
\usepackage{caption}
\usepackage{cmap}
\usepackage{multirow}
\usepackage{indentfirst}
\usepackage[usenames]{color}
\usepackage{amsthm}
\usepackage{xcolor}

\begin{document}

\title{Tridirectional quantum teleportation protocol in continuous variables}

\author{E. A. Nesterova}
\author{S. B. Korolev}
\affiliation{St.Petersburg State University, Universitetskaya nab. 7/9, St.Petersburg, 199034, Russia}
\begin{abstract}
    We propose a protocol for tridirectional quantum teleportation in continuous variables. A special feature of the protocol is the possibility to choose one of three scenarios: simultaneous exchange between three participants, exchange between any two participants, or the transfer of two states to a third participant. We use a cluster state in continuous variables as the main resource to realise tridirectional quantum teleportation. In the paper, we obtain several possible configurations of cluster states in continuous variables that can be used as the main resource. From the whole range of configurations, we have chosen those that realise the protocol with the smallest possible error.
\end{abstract}
\maketitle
\section{Introduction}
The quantum teleportation (QT) protocol was originally designed to transfer an unknown quantum state from one user (Alice) to another (Bob) \cite{Bennett}. However, this protocol has the limitation that teleportation is only possible in one direction, from Alice to Bob (referred to as "one-directional quantum teleportation". To exchange quantum information, Alice and Bob need to use either the QT protocol twice or another alternative protocol known as bidirectional quantum teleportation (BQT) \cite{Zha_BDQT,Fu2014,Li_GHZ,Yan2013,Duan2014,Chen2014,Nesterova_2024}. The BQT protocol allows two quantum states to be exchanged simultaneously. However, if we introduce a third participant, Charlie, the BQT protocol will not be sufficient. In this case, the tridirectional quantum teleportation (TQT) protocol becomes essential. The TQT protocol facilitates the exchange of quantum states between Alice, Bob, and Charlie \cite{Li2016,Singh2023}.

In the works \cite{Li2016,Singh2023}, a TQT protocol was proposed for discrete-variable physical systems.  This protocol uses multipartite entangled states as the main resource. However, entanglement operations in discrete-variable systems are probabilistic in nature \cite{Fu2014}. Using such operations leads to the fact that the probability of creating a large entangled state (with numerous connections) becomes tiny. The generation of a suitable resource is time-consuming, which has a detrimental effect on the rate of message transmission in the quantum network.

To solve this problem, one can use physical systems in continuous variables that are deterministically entangled \cite{Vaidman}. In \cite{Braunstein_telep_cv, Nesterova_2024}, it was shown that it is possible to implement QT and BQT of quantum states in continuous variables. This suggests that physical systems in continuous variables have the potential to be used for multiuser quantum teleportation. Our work will be devoted to the development of a TQT protocol in continuous variables.

However, it should be noted that continuous-variable physical systems also have drawbacks. To realize teleportation, the physical systems involved must be in squeezed states that are entangled with each other. The resulting multipartite entangled state forms the main resource. When such a resource is used, the teleported state will have an error that is proportional to the degree of squeezing of the states used, as well as to the number of physical systems involved \cite{Korolev_2020_1}.   The most evident approach to reducing the teleportation error in continuous variables is to use states with a higher squeezing degree. However, creating such states in experiments presents significant challenges \cite{Vahlbruch_2016}. Another method of reducing the error is to optimize the resource multipartite entangled state \cite{Korolev_2020}. The process of such optimization includes finding the optimal number of squeezed states, which are indispensable for the implementation of TQT, while concurrently minimizing the error. In addition, it is necessary to choose the correct configuration (entanglements between physical systems) of the resource state.

In our work, we will propose a TQT scheme in continuous variables. In this scheme, three users can exchange of their quantum states with each other. A special feature of the protocol is the possibility to choose one of several distinct scenarios. The protocol allows implementing three different TQT scenarios. The first scenario involves  a simultaneous exchange of quantum states between all participants. The second scenario describes a pairwise exchange of states for any two teleportation participants. And in the third, one of the participants receiving two quantum states from other users.

Furthermore, we will consider several ways of implementing the proposed TQT protocol in continuous variables, differing in the main resource. We will analyze the errors obtained when implementing the protocol in different ways, and optimize the protocol so that the error is minimally possible.

\section{Tridirectional teleportation protocol on a twelve-node cluster state in continuous variables}

\subsection{Generation of resource state} 

The goal of our work is to create a teleportation protocol in continuous variables that allows Alice, Bob, and Charlie to simultaneously transmit unknown states to each other. The initial stage is the generation of the main resource.  We will use a cluster state as this resource.  A cluster state is a quantum multipartite entangled state characterized by a mathematical graph. The nodes of the graph are physical systems in continuous variables, while the edges represent entanglements between them. 

The physical systems used to generate a cluster state are quantum oscillators in quadrature-squeezed states. Each such state is described by two observable operators $\hat{x}$ and $\hat{y}$. These operators obey the canonical commutation relation:
 \begin{align}
\left[ \hat{x}_j,\hat{y}_k\right]=\frac{i}{2}\delta _{j,k},
\end{align}
 where the indices $j$ and $k$ enumerate the corresponding oscillators and $\delta _{j,k}$ represents the Kronecker delta. We assume
 that all oscillators are squeezed in the $\hat y$-quadrature \cite{Gu}. This means that their variances are smaller than the variances
 of the vacuum state:
\begin{align} 
\langle \delta \hat{y}_j^2 \rangle < \frac{1}{4}, \qquad j=1,\dots,n.
\end{align}

To obtain the cluster state, we have to apply a Bogoliubov transformation \cite{Bogoljubov} to independent quantum oscillators. This transformation is given by the following matrix:
\begin{align} \label{U_Bogol}
U=(I+iA)(I+A^2)^{-1/2},
\end{align}
where $A$ is the adjacency matrix of the generated cluster state graph, and  $I$ is the identity matrix. The adjacency matrix $A$ is a symmetric matrix, that completely determines the cluster state graph. The elements of this matrix $A_{ij}$ are the weight coefficients of the cluster state graph. If the i-th and j-th nodes of the graph are not connected by an edge, the corresponding element of the adjacency matrix is set to zero.

\begin{figure}[h!]
    \centering
    \includegraphics[scale=1]{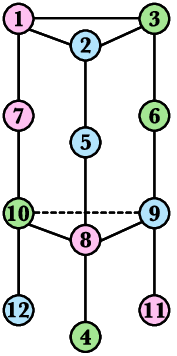}
    \caption{The twelve-node cluster state used for the TQT protocol in continuous variables. In the figure, nodes belonging to different participants are marked with different colors: pink indicates Alice's nodes, blue indicates Bob's nodes, and green indicates Charlie's nodes.}
    \label{fig:12_modes}
\end{figure}

To implement the TQT protocol, we will use a twelve-node cluster state, which has a graph shown in Fig.\ref{fig:12_modes}

The configuration of this cluster state (cluster state graph) was chosen based on symmetry for all three teleportation participants. Since the three teleportation participants should be completely equivalent, the distribution of cluster state nodes is required to be uniform. Consequently, the nodes: $1$, $7$, $8$, $11$ -- belong to Alice; $2$, $5$, $9$, $12$ -- to Bob; and $3$, $6$, $4$, $10$ -- to Charlie. The adjacency matrix of such a graph is given as follows:
\begin{align} \label{adj_matrix}
A_{12} = 
\begin{pmatrix}
0 & 1 & 1 & 0 & 0 & 0 & 1 & 0 & 0 & 0 & 0 & 0\\
1&0&1&0&1&0&0&0&0&0&0&0\\
1&1&0&0&0&1&0&0&0&0&0&0\\
0&0&0&0&0&0&0&1&0&0&0&0\\
0&1&0&0&0&0&0&1&0&0&0&0\\
0&0&1&0&0&0&0&0&1&0&0&0\\
1&0&0&0&0&0&0&0&0&1&0&0\\
0&0&0&1&1&0&0&0&-1&-1&0&0\\
0&0&0&0&0&1&0&-1&0&-1&1&0\\
0&0&0&0&0&0&1&-1&-1&0&0&1\\
0&0&0&0&0&0&0&0&1&0&0&0\\
0&0&0&0&0&0&0&0&0&1&0&0
\end{pmatrix}.
\end{align} 

Knowing the adjacency matrix, we can calculate the Bogoliubov transformation(\ref{U_Bogol}). Utilizing this transformation, the relationship between the quadrature operators of the cluster state and the quadrature operators of the independent squeezed quantum oscillators can be expressed as follows:
\begin{align}
\hat{\vec{X}}+i\hat{\vec{Y}}=\left(I+iA_{12}\right)\left(I+A_{12}^2\right)^{-1/2}\left( \hat{\vec{x}}_s+i\hat{\vec{y}}_s\right)\equiv\left(I+iA_{12}\right)\left( \hat{\vec{x}}_r+i\hat{\vec{y}}_r\right),
\end{align}
where $\hat{\vec{X}}=\left(\hat{X}_1,\hat{X}_2,\dots, \hat{X}_{12}\right)^T$ and $\hat{\vec{Y}}=\left(\hat{Y}_1,\hat{Y}_2,\dots, \hat{Y}_{12}\right)^T$ --are vectors consisting of quadratures of the cluster state; $\hat{\vec{x}}_s=\left(\hat{x}_{s,1},\hat{x}_{s,2},\dots, \hat{x}_{s,12}\right)^T$ and $\hat{\vec{y}}_s=\left(\hat{y}_{s,1},\hat{y}_{s,2},\dots, \hat{y}_{s,12}\right)^T$  are vectors consisting of quadratures of the squeezed quantum oscillators. To simplify the notation, we introduce new quadratures that are related to the quadratures of squeezed oscillators as follows: $\hat{\vec{x}}_r+i\hat{\vec{y}}_r=\left(I+A_{12}^2\right)\left( \hat{\vec{x}}_s+i\hat{\vec{y}}_s\right)$. Taking into account these reductions and the explicit form of the adjacency matrix (\ref{adj_matrix}), the vector of quadratures of the cluster state can be written as follows:
\begin{equation} \label{eq_0}
\begin{pmatrix}
\hat{X}_1+i\hat{Y}_1\\
\hat{X}_2+i\hat{Y}_2\\
\hat{X}_3+i\hat{Y}_3\\
\hat{X}_4+i\hat{Y}_4\\
\hat{X}_5+i\hat{Y}_5\\
\hat{X}_6+i\hat{Y}_6\\
\hat{X}_7+i\hat{Y}_7\\
\hat{X}_8+i\hat{Y}_8\\
\hat{X}_9+i\hat{Y}_9\\
\hat{X}_{10}+i\hat{Y}_{10}\\
\hat{X}_{11}+i\hat{Y}_{11}\\
\hat{X}_{12}+i\hat{Y}_{12}\\
\end{pmatrix}=
\begin{pmatrix}
\hat{x}_{r,1}-\hat{y}_{r,2}-\hat{y}_{r,3} -\hat{y}_{r,7} +i\left(\hat{y}_{r,1}+\hat{x}_{r,2}+\hat{x}_{r,3}+\hat{x}_{r,7}\right)\\
\hat{x}_{r,2} -\hat{y}_{r,1} -\hat{y}_{r,3} -\hat{y}_{r,5}+i(\hat{y}_{r,2}+\hat{x}_{r,1}+\hat{x}_{r,3}+\hat{x}_{r,5})\\
\hat{x}_{r,3} -\hat{y}_{r,1}-\hat{y}_{r,2} -\hat{y}_{r,6}+ i(\hat{y}_{r,3}+\hat{x}_{r,1}+\hat{x}_{r,2}+\hat{x}_{r,6})\\
\hat{x}_{r,4} -\hat{y}_{r,8}+i(\hat{y}_{r,4}+\hat{x}_{r,8})\\
\hat{x}_{r,5} -\hat{y}_{r,2}-\hat{y}_{r,8}+i(\hat{y}_{r,5}+\hat{x}_{r,2}+\hat{x}_{r,8})\\
\hat{x}_{r,6} -\hat{y}_{r,3}-\hat{y}_{r,9}+i(\hat{y}_{r,6}+\hat{x}_{r,3}+\hat{x}_{r,9})\\
\hat{x}_{r,7} -\hat{y}_{r,1}-\hat{y}_{r,10}+i(\hat{y}_{r,7}+\hat{x}_{r,1}+\hat{x}_{r,10})\\
\hat{x}_{r,8} -\hat{y}_{r,4} -\hat{y}_{r,5} +\hat{y}_{r,9}+\hat{y}_{r,10}+i(\hat{y}_{r,8}+\hat{x}_{r,4}+\hat{x}_{r,5}-\hat{x}_{r,9}-\hat{x}_{r,10})\\
\hat{x}_{r,9} -\hat{y}_{r,6} -\hat{y}_{r,11} +\hat{y}_{r,8}+\hat{y}_{r,10}+i(\hat{y}_{r,9}+\hat{x}_{r,6}+\hat{x}_{r,11}-\hat{x}_{r,8}-\hat{x}_{r,10})\\
\hat{x}_{r,10} -\hat{y}_{r,7} -\hat{y}_{r,12} +\hat{y}_{r,8}+\hat{y}_{r,9}+i(\hat{y}_{r,10}+\hat{x}_{r,7}+\hat{x}_{r,12}-\hat{x}_{r,8}-\hat{x}_{r,9})\\
\hat{x}_{r,11} -\hat{y}_{r,9}+i(\hat{y}_{r,11}+\hat{x}_{r,9})\\
\hat{x}_{r,12} -\hat{y}_{r,10}+i(\hat{y}_{r,12}+\hat{x}_{r,10})\\
\end{pmatrix}.
\end{equation}
Once we have obtained the cluster state, we can proceed with the TQT procedure.

\subsection{Tridirectional teleportation: Bob -- Alice -- Charlie -- Bob}
\begin{figure}[h!]
    \centering
    \includegraphics[scale=0.9]{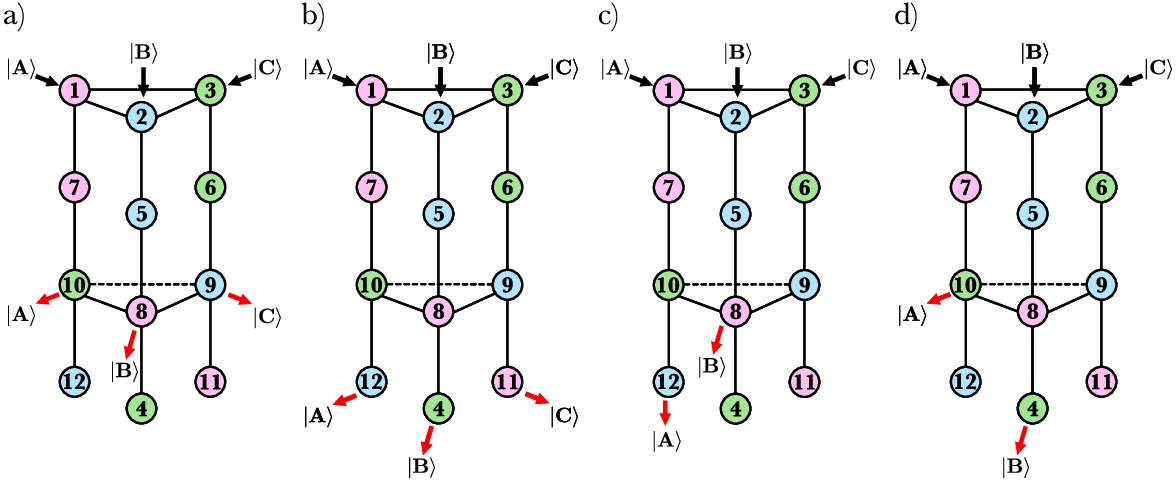}
    \caption{
    Different implementations of the tridirectional teleportation protocol: a) Alice transmits to Charlie, Charlie transmits to Bob, and Bob transmits to Alice; b) Alice transmits to Bob, Bob transmits to Charlie, and Charlie transmits to Alice; c) Alice and Bob exchange states; d) Alice and Bob transmit their states to Charlie. In the figure, nodes belonging to different participants are marked with different colors: pink indicates Alice's nodes, blue indicates Bob's nodes, and green indicates Charlie's nodes.}
    \label{fig:cases}
\end{figure}

The present study commences with an examination of the TQT in the scenario where three participants (Bob, Alice, and Charlie) exchange their states circularly. Bob passes his state, described by the annihilation operator $\hat{b}_{in}=\hat{x}_b+i\hat{y}_b$, to Alice. In turn, Alice passes her state ($\hat{a}_{in}=\hat{x}_a+i\hat{y}_a$) to Charlie, and Charlie passes his state ($\hat{c}_{in}=\hat{x}_c+i\hat{y}_c$) to Bob. This case is depicted in Fig. \ref{fig:cases} a).

 For teleportation, Alice, Bob, and Charlie entangle their teleported states into a common cluster state that is shared between them. To entangle their teleported states, Alice and Bob use symmetric beam splitters whose transformation matrix has the form:
\begin{align}
    U_{BS}=\frac{1}{\sqrt{2}}\begin{pmatrix}
        1 & 1\\
        1 & -1
    \end{pmatrix}.
\end{align}
 Alice entangles her teleported state with the state in the first node of the cluster state, Bob entangles his teleported state with the state in the second node, and Charlie entangles his teleported state with the state in the third node. Thereafter, the quadratures of Alice's, Bob's, and Charlie's input modes, as well as the quadratures in the first, second, and third modes of the cluster state, are transformed as follows:
\begin{align} \label{13}
\hat{A}'_{in}=\frac{1}{\sqrt{2}}(\hat{x}_a +\hat{x}_{r,1} -\hat{y}_{r,2} - \hat{y}_{r,3}-\hat{y}_{r,7}+i(\hat{y}_a+\hat{y}_{r,1}+\hat{x}_{r,2}+\hat{x}_{r,3}+\hat{x}_{r,7})),\\
\hat{A}'_{1}=\frac{1}{\sqrt{2}}(\hat{x}_a -\hat{x}_{r,1} +\hat{y}_{r,2} + \hat{y}_{r,3}+\hat{y}_{r,7}+i(\hat{y}_a-\hat{y}_{r,1}-\hat{x}_{r,2}-\hat{x}_{r,3}-\hat{x}_{r,7})),\\
\hat{B}'_{in}=\frac{1}{\sqrt{2}}(\hat{x}_b +\hat{x}_{r,2} -\hat{y}_{r,1} - \hat{y}_{r,3}-\hat{y}_{r,5}+i(\hat{y}_b+\hat{y}_{r,2}+\hat{x}_{r,1}+\hat{x}_{r,3}+\hat{x}_{r,5})),\\
\hat{B}'_{1}=\frac{1}{\sqrt{2}}(\hat{x}_b -\hat{x}_{r,2} +\hat{y}_{r,1} + \hat{y}_{r,3}+\hat{y}_{r,5}+i(\hat{y}_b-\hat{y}_{r,2}-\hat{x}_{r,1}-\hat{x}_{r,3}-\hat{x}_{r,5})),\\
\hat{C}'_{in}=\frac{1}{\sqrt{2}}(\hat{x}_c +\hat{x}_{r,3} -\hat{y}_{r,1} - \hat{y}_{r,2}-\hat{y}_{r,6}+i(\hat{y}_c+\hat{y}_{r,3}+\hat{x}_{r,1}+\hat{x}_{r,2}+\hat{x}_{r,6})),\\
\hat{C}'_{1}=\frac{1}{\sqrt{2}}(\hat{x}_c -\hat{x}_{r,3} +\hat{y}_{r,1} + \hat{y}_{r,2}+\hat{y}_{r,6}+i(\hat{y}_c-\hat{y}_{r,3}-\hat{x}_{r,1}-\hat{x}_{r,2}-\hat{x}_{r,6})).
\end{align}

The next step in the teleportation procedure is measurement. Alice, Bob, and Charlie measure the fields obtained at the output of the beam splitter, and the fields in modes $5$-$7$, $4$, and $11$-$12$. The unmeasured modes, in this case $8$, $9$, $10$, are the output nodes (their states are the result of teleportation). For the measurements, we use homodyne detectors that measure a superposition of quadratures of the field. The amplitude of the quantum photocurrent obtained as a result of homodyne measurement of the field with quadratures $\hat{x}$ and $\hat{y}$ can be written as follows:
 \begin{align} 
 \hat{i}=\beta _0 \left(\cos \theta \hat{x}+\sin \theta \hat{y}\right),
 \end{align}
 where $\beta _0$  is the amplitude of the local oscillator, and ${\theta}$  is its phase. The quantum photocurrents obtained as a result of all measurements can be written as the following system of equations:
\begin{equation} \label{eq_cases}
    \begin{cases}
        \sin \theta_{A_{in}} \hat{Y}_{A_{in}'}+\cos \theta_{A_{in}} \hat{X}_{A_{in}'}=\frac{\hat{i}_{A_{in}}}{\beta_0},\\
        
   \sin \theta_{A_1} \hat{Y}_{A_1'}+\cos \theta_{A_1} \hat{X}_{A_1'}=\frac{\hat{i}_{A_1}}{\beta_0},\\
   \sin \theta_{B_{in}} \hat{Y}_{B_{in}'}+\cos \theta_{B_{in}} \hat{X}_{B_{in}'}=\frac{\hat{i}_{B_{in}}}{\beta_0},\\
        
   \sin \theta_{B_1} \hat{Y}_{B_1'}+\cos \theta_{B_1} \hat{X}_{B_1'}=\frac{\hat{i}_{B_1}}{\beta_0},\\
   \sin \theta_{C_{in}} \hat{Y}_{C_{in}'}+\cos \theta_{C_{in}} \hat{X}_{C_{in}'}=\frac{\hat{i}_{C_{in}}}{\beta_0},\\
        
   \sin \theta_{C_1} \hat{Y}_{C_1'}+\cos \theta_{C_1} \hat{X}_{C_1'}=\frac{\hat{i}_{C_1}}{\beta_0},\\

    \sin \theta_4 \hat{Y}_4+\cos \theta_4 \hat{X}_4=\frac{\hat{i}_4}{\beta_0},\\
    
    \sin \theta_5 \hat{Y}_5+\cos \theta_5 \hat{X}_5=\frac{\hat{i}_5}{\beta_0},\\
    
    \sin \theta_6 \hat{Y}_6+\cos \theta_6 \hat{X}_6=\frac{\hat{i}_6}{\beta_0},\\
    
    \sin \theta_7 \hat{Y}_7+\cos \theta_7 \hat{X}_7=\frac{\hat{i}_7}{\beta_0},\\
    
    \sin \theta_{11} \hat{Y}_{11}+\cos \theta_{11} \hat{X}_{11}=\frac{\hat{i}_{11}}{\beta_0},\\
    \sin \theta_{12} \hat{Y}_{12}+\cos \theta_{12} \hat{X}_{12}=\frac{\hat{i}_{12}}{\beta_0},\\
 \end{cases}      
\end{equation}

The modes of the system under study are entangled, so measuring some modes will affect others. We solve the system of Eqs. (\ref{eq_cases}) for the quadratures $\hat{x}_{r,i}$ to track this influence and substitute this solution into the expressions for the unmeasured quadratures. We analyze the relation between the unmeasured quadratures and the input states. As a result, we can find the phases of the homodyne detectors that lead to the implementation of the TQT protocol. These phases are as follows:
\begin{align}
    &\theta_{A_{in}}=\theta_{B_{in}}=\theta_{C_{in}}=0,\\
    &\theta_{A_{1}}=\theta_{B_{1}}=\theta_{C_{1}}=\frac{\pi}{2},\\
    &\theta_5=\theta_6=\theta_7=\frac{\pi}{2},\\
     &\theta_4=\theta_{11}=\theta_{12}=0.
\end{align}
Taking into account the obtained restrictions on the phases of homodyne detectors, the relation between the output (unmeasured) and input quadratures can be written in vector form as follows: 
\begin{align} \label{out_twent}
    \begin{pmatrix}
        \hat{X}_{out,A}\\
        \hat{X}_{out,B}\\
        \hat{X}_{out,C}\\
        \hat{Y}_{out,A}\\
        \hat{Y}_{out,B}\\
        \hat{Y}_{out,C}
    \end{pmatrix}\equiv \begin{pmatrix}
        \hat{X}_{8}\\
        \hat{X}_{9}\\
        \hat{X}_{10}\\
        \hat{Y}_{8}\\
        \hat{Y}_{9}\\
        \hat{Y}_{10}
    \end{pmatrix}=\begin{pmatrix}
        \hat{x}_{b}\\
        \hat{x}_c\\
        \hat{x}_a\\
        \hat{y}_b\\
        \hat{y}_c\\
        \hat{y}_a
    \end{pmatrix}+\frac{1}{\beta_0}\begin{pmatrix}
        \hat{i}_5-\sqrt{2}\hat{i}_{B_{in}}\\
        \hat{i}_6-\sqrt{2}\hat{i}_{C_{in}}\\
        \hat{i}_7-\sqrt{2}\hat{i}_{A_{in}}\\
        \hat{i}_4-\hat{i}_6-\hat{i}_7-\sqrt{2}\hat{i}_{B_{1}}\\
        \hat{i}_{11}-\hat{i}_5-\hat{i}_7-\sqrt{2}\hat{i}_{C_{1}}\\
        \hat{i}_{12}-\hat{i}_5-\hat{i}_6-\sqrt{2}\hat{i}_{A_{1}}\\
    \end{pmatrix}+\hat{\vec{E}}_{BCA},
\end{align}
where, to shorten the notation, a vector was introduced:
\begin{align} \label{vec_err_bca}
\hat{\vec{E}}_{BCA}=\left(\begin{array}{cccccccccccc}
-1&0&-1&-1&-3&0&0&0&1&1&0&0\\
-1&-1&0&0&0&-3&0&1&0&1&-1&0\\
0&-1&-1&0&0&0&-3&1&1&0&0&-1\\
0&-1&0&0&0&1&1&2&0&0&0&0\\
0&0&-1&0&1&0&1&0&2&0&0&0\\
-1&0&0&0&1&1&0&0&0&2&0&0
    \end{array}\right)\begin{pmatrix}
        \hat{y}_{r,1}\\
        \hat{y}_{r,2}\\
        \hat{y}_{r,3}\\
        \hat{y}_{r,4}\\
        \hat{y}_{r,5}\\
        \hat{y}_{r,6}\\
        \hat{y}_{r,7}\\
        \hat{y}_{r,8}\\
        \hat{y}_{r,9}\\
        \hat{y}_{r,10}\\
        \hat{y}_{r,11}\\
        \hat{y}_{r,12}\\
    \end{pmatrix}.
\end{align}

In Eq (\ref{out_twent}), the first term on the right-hand side indicates that the TQT has been successful. Alice, Bob, and Charlie have exchanged their states. 

The second term on the right-hand side of Eq. (\ref{out_twent}) corresponds to the measured photocurrents. Once all the measurements have been completed, the photocurrent operators are replaced with the real numbers obtained from the measurements. Since Alice, Bob, and Charlie have access to their photocurrent values from the experiment, they can communicate these values to each other and adjust all the classical quantities (real terms) in the output quadratures using a quadrature displacement operation. As a result of this displacement, the second term in the vector from Eq. (\ref{out_twent}) will be eliminated, meaning that the teleportation results will no longer depend on the measurement outcomes.

The third term on the right-hand side of the Eq. (\ref{out_twent}) is the error. This error is proportional to the combinations of squeezed quadratures of the quantum oscillators used. To estimate the magnitude of errors, it is necessary to move from error operators to their variances. In doing so, we take into account that the squeezed oscillators used are independent and have the same squeezing degree. I.e., the following relation is valid:
$\langle \delta \hat{y}_{s,i}\delta \hat{y}_{s,j} \rangle=\delta_{ij}\langle \delta \hat{y}_s^2 \rangle$, where $\delta_{ij}$ -- is the Kronecker delta. Considering this along with the fact that $\hat{\vec{y}}_r=\left(I+A_{12}^2\right)\hat{\vec{y}}_s$, we can rewrite the error vector (\ref{vec_err_bca}) as a vector of its error variances:
\begin{align} \label{err_BCA}
    \delta\hat{\vec{e}}^2_{BCA}=\begin{pmatrix}
        5\\
        5\\
        5\\
       3\\
        3\\
         3
    \end{pmatrix}  \langle \delta \hat{y}_s^2 \rangle.
\end{align}
We see that this vector is proportional to the variances of the squeezed quadrature of the oscillators used. It is evident that the greater the degree of squeezing of the quantum oscillators employed, the smaller the error variance will be, and therefore the less the errors will affect the results of TQT. Below, we will discuss the teleportation errors will be examined in more detail, and the protocol will be optimized so that the error is minimal.

\subsection{Tridirectional teleportation: Alice -- Bob -- Charlie -- Alice}
Now consider another scenario of TQT, in which the transmission is in the opposite direction. Alice transmits her state to Bob, Bob transmits his state to Charlie, and Charlie sends his state to Alice. This case is shown in Fig. \ref{fig:cases} b).

To teleport, Alice, Bob, and Charlie entangle their input states with the first, second, and third nodes of the cluster state, respectively, using symmetric beam splitters. Then, they perform homodyne measurements of all states except those at nodes $11$, $12$, and $4$ -- the output nodes. Finally, knowing the results of each other's homodyne measurements, they compensate the classical fields and obtain the teleportation results.

Let us analyze the equations describing this case. We have found that for successful teleportation it is necessary to select the following phases of local oscillators of homodyne detectors:
    \begin{align}
    &\theta_{A_{in}}=\theta_{B_{in}}=\theta_{C_{in}}=\frac{\pi}{4},\\
    &\theta_{A_{1}}=\theta_{B_{1}}=\theta_{C_{1}}=\frac{3\pi}{4},\\
    &\theta_5=\theta_6=\theta_7=\frac{\pi}{2},\\
     &\theta_8=\theta_{9}=\theta_{10}=\frac{\pi}{2}.
\end{align}
the relation between the output and input quadratures can be written in vector form as follows:
\begin{align} 
    \begin{pmatrix}
        \hat{X}_{out,A}\\
        \hat{X}_{out,B}\\
        \hat{X}_{out,C}\\
        \hat{Y}_{out,A}\\
        \hat{Y}_{out,B}\\
        \hat{Y}_{out,C}
    \end{pmatrix}\equiv \begin{pmatrix}
        \hat{X}_{11}\\
        \hat{X}_{12}\\
        \hat{X}_{4}\\
        \hat{Y}_{11}\\
        \hat{Y}_{12}\\
        \hat{Y}_4
    \end{pmatrix}=\begin{pmatrix}
        \hat{x}_c\\
        \hat{x}_a\\
        \hat{x}_b\\
        \hat{y}_{c}\\
        \hat{y}_a\\
        \hat{y}_b
    \end{pmatrix}+\hat{\vec{E}}_{CAB},
\end{align}
where, to shorten the notation, a vector was introduced:
\begin{align}
\hat{\vec{E}}_{CAB}=\begin{pmatrix}
0&0&1&0&-1&0&-1&0&-2&0&0&0\\
1&0&0&0&-1&-1&0&0&0&-2&0&0\\
0&1&0&0&0&-1&-1&-2&0&0&0&0\\
-1&-1&0&0&0&-2&0&0&0&0&1&0\\
0&-1&-1&0&0&0&-2&0&0&0&0&1\\
-1&0&-1&1&-2&0&0&0&0&0&0&0
\end{pmatrix}\begin{pmatrix}
        \hat{y}_{r,1}\\
        \hat{y}_{r,2}\\
        \hat{y}_{r,3}\\
        \hat{y}_{r,4}\\
        \hat{y}_{r,5}\\
        \hat{y}_{r,6}\\
        \hat{y}_{r,7}\\
        \hat{y}_{r,8}\\
        \hat{y}_{r,9}\\
        \hat{y}_{r,10}\\
        \hat{y}_{r,11}\\
        \hat{y}_{r,12}\\
    \end{pmatrix}.
    \end{align}
 As in the previous case, we obtain TQT with the error. The values of error variances can be written in vector form as follows:   
\begin{align} \label{err_CAB}
    \delta\hat{\vec{e}}^2_{CAB}=\begin{pmatrix}
        3\\
        3\\
        3\\
       5\\
        5\\
         5
    \end{pmatrix}   \langle \delta \hat{y}_s^2 \rangle.
\end{align}
As was the case previously, it is evident that an increase in squeezing degree results in a decrease in teleportation error.

\subsection{Bidirectional teleportation with the common cluster state}
In the previous subsections, we have considered two possible cases of TQT involving three participants. However, in practice, a scenario may emerge wherein one of the participants (Charlie) is not required to teleport any information. The exchange of states is conducted exclusively between Alice and Bob (BQT).  The implementation of this protocol is shown in Fig. \ref{fig:cases} c).

To implement such a protocol, Alice and Bob entangle their teleported states with symmetric beam splitters to the first and second nodes of the cluster state, respectively. Then they perform homodyne measurements. In this case, Alice's and Bob's output nodes are nodes with numbers $8$ and $12$, respectively. The phases of the local oscillators of the homodyne detectors must be chosen as follows:
 \begin{align}
    &\theta_{A_{in}}=\frac{\pi}{4},\\
    &\theta_{B_{in}}=0,\\
     &\theta_{A_{1}}=\frac{3\pi}{4},\\
      &\theta_{B_{1}}=\frac{\pi}{2},\\
    &\theta_5=\theta_6=\theta_7=\theta_{10}=\frac{\pi}{2},\\
     &\theta_4=0.
\end{align}
Alice and Bob will be able to exchange quantum states with each other. In this case, the output quadrature vector can be written as follows:
\begin{align} 
    \begin{pmatrix}
        \hat{X}_{out,A}\\
        \hat{X}_{out,B}\\
        \hat{Y}_{out,A}\\
        \hat{Y}_{out,B}
    \end{pmatrix}\equiv \begin{pmatrix}
        \hat{X}_{8}\\
        \hat{X}_{12}\\
        \hat{Y}_{8}\\
        \hat{Y}_{12}
    \end{pmatrix}=\begin{pmatrix}
        \hat{x}_{b}\\
        \hat{x}_a\\
        \hat{y}_b\\
        \hat{y}_a
    \end{pmatrix}+\hat{\vec{E}}_{BA},
\end{align}
where
\begin{align}
\hat{\vec{E}}_{BA}=\left(\begin{array}{cccccccccccc}
-1&0&-1&-1&-3&0&0&0&1&1&0&0\\
1&0&0&0&-1&-1&0&0&0&-2&0&0\\
0&-1&0&0&0&1&1&2&0&0&0&0\\
0&-1&-1&0&0&0&-2&0&0&0&0&1
    \end{array}\right)\begin{pmatrix}
        \hat{y}_{r,1}\\
        \hat{y}_{r,2}\\
        \hat{y}_{r,3}\\
        \hat{y}_{r,4}\\
        \hat{y}_{r,5}\\
        \hat{y}_{r,6}\\
        \hat{y}_{r,7}\\
        \hat{y}_{r,8}\\
        \hat{y}_{r,9}\\
        \hat{y}_{r,10}\\
        \hat{y}_{r,11}\\
        \hat{y}_{r,12}\\
    \end{pmatrix}.
    \end{align}
The vector of error variances can be written as follows:
\begin{align} \label{err_BA}
    \delta\hat{\vec{e}}^2_{BA}=\begin{pmatrix}
        3\\
        5\\
        5\\
       3
    \end{pmatrix}   \langle \delta \hat{y}_s^2 \rangle.
\end{align}

It is important to note here that such a case of teleportation does not depend on whether the states in nodes $3$, $9$, and $11$ were measured. The states in these nodes without measurement remain entangled and do not affect the implementation of the BQT procedure. 

A protocol for exchanging states between Alice and Charlie or Charlie and Bob can be implemented similarly.

\subsection{Teleportation of two states to the third participant}
Such a cluster state can be used for more than just exchanging quantum states. Let us consider the TQT case, where Alice and Bob transmit their states to Charlie.  This scenario is illustrated in Fig. \ref{fig:cases} d).

To implement such a protocol, Alice and Bob, as before, entangle their teleported states with the first and second nodes. Then they perform homodyne measurements. Selecting the phases of local oscillators as follows:
\begin{align}
    &\theta_{A_{in}}=\pi,\\
    &\theta_{B_{in}}=\frac{\pi}{4},\\
     &\theta_{A_{1}}=\theta_3=\theta_5=\theta_6=\theta_{7}=\theta_8=\frac{\pi}{2},\\
      &\theta_{B_{1}}=\frac{3\pi}{4},\\
     &\theta_9=\theta_{11}=\theta_{12}=0,
\end{align}
we can teleport the states of Alice and Bob to nodes belonging to Charlie. In this case, the output quadrature vector will look like:
\begin{align} 
    \begin{pmatrix}
        \hat{X}_{out,C_a}\\
        \hat{X}_{out,C_b}\\
        \hat{Y}_{out,C_a}\\
        \hat{Y}_{out,C_b}
    \end{pmatrix}\equiv \begin{pmatrix}
        \hat{X}_{10}\\
        \hat{X}_{4}\\
        \hat{Y}_{10}\\
        \hat{Y}_{4}
    \end{pmatrix}=\begin{pmatrix}
        \hat{x}_{a}\\
        \hat{x}_b\\
        \hat{y}_a\\
        \hat{y}_b
    \end{pmatrix}+\hat{\vec{E}}_{CC},
\end{align}
where
\begin{align}
\hat{\vec{E}}_{CC}=\left(\begin{array}{cccccccccccc}
0&-1&-1&0&0&0&-3&1&1&0&0&-1\\
0&1&0&0&0&-1&-1&-2&0&0&0&0\\
-1&0&0&0&1&1&0&0&0&2&0&0\\
-1&0&-1&1&-2&0&0&0&0&0&0&0
    \end{array}\right)\begin{pmatrix}
        \hat{y}_{r,1}\\
        \hat{y}_{r,2}\\
        \hat{y}_{r,3}\\
        \hat{y}_{r,4}\\
        \hat{y}_{r,5}\\
        \hat{y}_{r,6}\\
        \hat{y}_{r,7}\\
        \hat{y}_{r,8}\\
        \hat{y}_{r,9}\\
        \hat{y}_{r,10}\\
        \hat{y}_{r,11}\\
        \hat{y}_{r,12}\\
    \end{pmatrix}.
    \end{align}
The vector of error variances in this case completely coincides with the case of BQT described in the previous section.
\begin{align} \label{err_cc}
    \delta\hat{\vec{e}}^2_{CC}=\begin{pmatrix}
        3\\
        5\\
        5\\
       3
    \end{pmatrix}   \langle \delta \hat{y}_s^2 \rangle.
\end{align}

\section{Tridirectional teleportation protocol on three three-mode cluster states}
\subsection{Tridirectional teleportation protocol with three independent cluster states}
\begin{figure}[h!]
    \centering
    \includegraphics[scale=0.9]{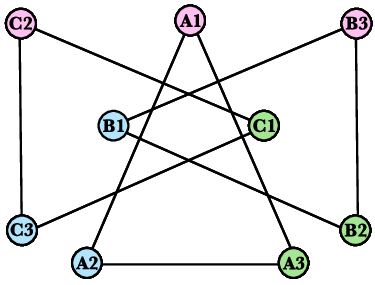}
    \caption{Three independent three-node cluster states used for tridirectional teleportation. In the figure, nodes belonging to different participants are marked with different colors: pink indicates Alice’s nodes, blue indicates Bob’s nodes, and green indicates Charlie’s nodes.}
    \label{fig:three_clust}
\end{figure}

As we have already established, TQT in continuous variables is subject to an inherent error. This error is proportional to the squeezing of the used states and the number of oscillators employed. To minimize the teleportation error, it is necessary to reduce the number of nodes used to implement the TQT. As the main resource, three independent three-node cluster states will be utilized, which are distributed equally among all participants of the teleportation protocol, as shown in Fig. \ref{fig:three_clust}. 

Each node of the three-node cluster state is distributed among the users. Each user entangles their input state with one of the nodes, while the remaining two nodes are distributed among the other participants. After distribution, Alice owns nodes $A1, B3$, and $C2$, Bob owns nodes $B1, A2$, and $C3$, and Charlie owns nodes $C1, A3$, and $B2$. This configuration ensures complete equivalence among the users. Notably, this setup reduces the number of squeezed oscillators required for teleportation from twelve to nine.

Due to the fact that each three-node cluster state is independent of the other, the teleportation of quantum states will also occur independently of the others. We will prove the possibility of implementing TQT in the scheme under consideration. To do this, it is enough to demonstrate that, by selecting the phases of local oscillators, it is possible to teleport to different nodes of a three-node linear cluster state.

\subsection{Teleportation to different nodes of a three-node cluster state}
\begin{figure}[h!]
    \centering
    \includegraphics[scale=1.0]{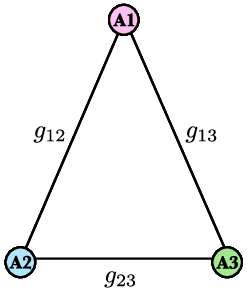}
    \caption{Weighted graph of three-node cluster state.}
    \label{fig:3mode_cluster}
\end{figure}

Let us consider teleportation to different nodes of a three-node cluster state. As the main resource, we will use the cluster state with the graph shown in Fig.\ref{fig:3mode_cluster}.

The ability to generate cluster states with different weight coefficients is one of the advantages of using physical systems in continuous variables. The weight coefficients of cluster states provide an additional degree of freedom, which can be optimized to enhance quantum information protocols based on cluster states \cite{Korolev_2018,Zinatullin_2022}. Furthermore, it has been demonstrated in \cite{Nesterova_2024} that by carefully selecting these weight coefficients, the errors in BQT can be reduced. With this in mind, our goal in this section is to determine the optimal weight coefficients for a three-node cluster state. These coefficients should enable teleportation to different nodes of the cluster while minimizing teleportation errors.

As before, we will use the adjacency matrix to describe the graph of the weighted cluster state. In the general case, for a three-node weighted cluster state, the adjacency matrix will have the following form:
\begin{equation}
A_{3} = \left(
\begin{array}{ccc}
0&g_{12}&g_{13}\\
g_{12}&0&g_{23}\\
g_{13}&g_{23}&0\\
\end{array}
\right).
\end{equation}

Knowing the adjacency matrix, we can calculate the matrix of the Bogoliubov transformation. Using the notation $\hat{\vec{x}}_r+i\hat{\vec{y}}_r=\left(I+A_{3}^2\right)\left( \hat{\vec{x}}_s+i\hat{\vec{y}}_s\right)$, we write the relation between the quadratures of the cluster state and the linear combinations of the quadratures of the squeezed states:
\begin{equation} \label{eq_0}
\begin{pmatrix}
\hat{X}_1+i\hat{Y}_1\\
\hat{X}_2+i\hat{Y}_2\\
\hat{X}_3+i\hat{Y}_3\\
\end{pmatrix}=
\begin{pmatrix}
\hat{x}_{r,1} -g_{12}\hat{y}_{r,2}-g_{13}\hat{y}_{r,3}+i(\hat{y}_{r,1}+g_{12}\hat{x}_{r,2}+g_{13}\hat{x}_{r,3})\\
\hat{x}_{r,2} -g_{12}\hat{y}_{r,1}-g_{23}\hat{y}_{r,3}+i(\hat{y}_{r,2}+g_{12}\hat{x}_{r,1}+g_{23}\hat{x}_{r,3})\\
\hat{x}_{r,3} -g_{13}\hat{y}_{r,1}-g_{23}\hat{y}_{r,2}+i(\hat{y}_{r,3}+g_{13}\hat{x}_{r,1}+g_{23}\hat{x}_{r,2})\\
\end{pmatrix}.
\end{equation}
After we have completed the generation of the main resource, we need to distribute the nodes of the cluster state among the participants. Let Alice own the central node $A1$. Nodes $A2$ and $A3$ belong to Bob and Charlie, respectively.

As the first step in the teleportation protocol, Alice entangles her teleported state $\hat{a}_{in}=\hat{x}_a+i\hat{y}_a$ into her node $A1$ using a symmetric beam splitter. As a result of this entanglement, the quadrature of Alice's input mode and the quadrature of the field, are transformed as follows:
\begin{align} \label{}
 &\hat{X}_{{in}}'+i\hat{Y}_{{in}}'= \frac{1}{\sqrt{2}}(\hat{x}_{a}+\hat{x}_{r,1} -g_{12} \hat{y}_{r,2} - g_{13} \hat{y}_{r,3}+i(\hat{y}_{a}+\hat{y}_{r,1} +g_{12} \hat{x}_{r,2} + g_{13} \hat{x}_{r,3}),\\
 &\hat{X}_{a_{1}}+i\hat{Y}_{a_{1}}= \frac{1}{\sqrt{2}}(\hat{x}_{a}-\hat{x}_{r,1} +g_{12} \hat{y}_{r,2} + g_{13} \hat{y}_{r,3}+i(\hat{y}_{a}-\hat{y}_{r,1} -g_{12}\hat{x}_{r,2} -g_{13}\hat{x}_{r,3})).
 \end{align}

If Alice plans to teleport her state to node A3 (Charlie), then it is necessary to measure the fields at the output of the beam splitter and the field in mode A2, choosing the phases of the local oscillators as follows:
\begin{align}
&\tan\theta_{in}=-g_{13},\\
&\tan \theta_{\text{A1}} = g_{13},\\
&\tan \theta_{\text{A2}} = 0.   
\end{align} 

As a result of such teleportation, the relation between the output and input quadratures can be written as follows:
\begin{align} \label{out_5}
    \begin{pmatrix}
        \hat{X}_{out}\\
        \hat{Y}_{out}
    \end{pmatrix}\equiv \begin{pmatrix}
        \hat{X}_{\text{A3}}\\
        \hat{Y}_{\text{A3}}
    \end{pmatrix}=\begin{pmatrix}
        \hat{x}_{a}\\
        \hat{y}_a
    \end{pmatrix}+\hat{\vec{e}}_1,
\end{align}
where
\begin{align} \label{error_vec}
   \hat{\vec{e}}_1= \begin{pmatrix}
        -\frac{1}{g_{13}} - \frac{g_{12}^2}{g_{13}} - g_{13}&&-g_{23}&&-\frac{g_{12}g_{23}}{g_{13}}\\
        g_{12} g_{23}&&g_{12}g_{13}&&1 + g_{13}^2 + g_{23}^2\\
    \end{pmatrix}\begin{pmatrix}
        \hat{y}_{r,1}\\
        \hat{y}_{r,2}\\
        \hat{y}_{r,3}
    \end{pmatrix}.
\end{align}
Here are the expressions after the quadrature displacement based on the measured values.

The case where Alice plans to teleport her state to node A2 (Bob) is symmetric to the previous one up to substitutions of $g_{13}\rightarrow g_{12}$ and $g_{12}\rightarrow g_{13}$. Thus, performing homodyne measurements with phases:
\begin{align}
&\tan\theta_{in}=-g_{12},\\
&\tan \theta_{\text{A1}} = g_{12},\\
&\tan \theta_{\text{A3}} = 0,   
\end{align} 
we get the following relation between the input and output quadratures
\begin{align}
    \begin{pmatrix}
        \hat{X}_{out}\\
        \hat{Y}_{out}
    \end{pmatrix}\equiv \begin{pmatrix}
        \hat{X}_{\text{A2}}\\
        \hat{Y}_{\text{A2}}
    \end{pmatrix}=\begin{pmatrix}
        \hat{x}_{a}\\
        \hat{y}_a
    \end{pmatrix}+\hat{\vec{e}}_2,
\end{align}
where
\begin{align} \label{error_vec_2}
   \hat{\vec{e}}_2= \begin{pmatrix}
        -\frac{1}{g_{12}} - \frac{g_{13}^2}{g_{12}} - g_{12}&&-g_{23}&&-\frac{g_{13}g_{23}}{g_{12}}\\
        g_{13} g_{23}&&g_{13}g_{12}&&1 + g_{12}^2 + g_{23}^2\\
    \end{pmatrix}\begin{pmatrix}
        \hat{y}_{r,1}\\
        \hat{y}_{r,2}\\
        \hat{y}_{r,3}
    \end{pmatrix}.
\end{align}

Let us analyze the error vectors (\ref{error_vec}) and (\ref{error_vec_2}). Note that the teleportation error is determined by the vector of squeezed $\hat{y}$-quadratures of the quantum oscillators and a matrix that depends solely on the weight coefficients. We can reduce the values of this matrix by setting $g_{23}$=0. Furthermore, the TQT protocol will become symmetrical, and the participants will be fully equivalent if we set $g_{12}=g_{13}=g$. In this case, the teleportation error will be uniform, and the vector of error variances will take the following form:
\begin{align} \label{vec_kor}
    \langle \delta \hat{\vec{e}}^2_1 \rangle=\begin{pmatrix}
    2+ \frac{1}{g^2}\\
    1+g^2
    \end{pmatrix} \langle \delta \hat{y}_s^2 \rangle,
\end{align}

Since all three-node cluster states are identical, when implementing TQT (simultaneous teleportation on each of the three-node clusters), the vector of error variances will be constructed from the vectors (\ref{vec_kor}) and will take the following form:
\begin{align}  \label{vec_3_tel}
  \langle \delta \hat{\vec{e}}^2 \rangle=  \begin{pmatrix}
        2+ \frac{1}{g^2}\\
        2+ \frac{1}{g^2}\\
        2+ \frac{1}{g^2}\\
        1+g^2\\
        1+g^2\\
         1+g^2
    \end{pmatrix}  \langle \delta \hat{y}_s^2 \rangle.
\end{align}
Here is a vector for a specific TQT case (Bob -- Alice -- Charlie -- Bob). Depending on the TQT scenario being implemented, the vectors of error variances will differ only by the permutation of the elements. The magnitude of the vector will depend only on the weight coefficient $g$.

Let us compare the vectors of error variances (\ref{vec_3_tel}) with the vector (\ref{err_BCA}). It is evident that for $1/\sqrt{3}<g<\sqrt{2}$, the vector of error variances obtained with this implementation of the TQT protocol is smaller than the vector obtained using a twelve-node cluster state. Thus, by utilizing three independent three-node cluster states, we can implement TQT with the same functionality as before but with reduced error.

\section{Tridirectional teleportation protocol on a set of two-node cluster states}
Consider another way of implementing TQT. As the main resource, we use independent two-node cluster states. In \cite{Nesterova_2024} the implementation of BQT using two-node cluster states was investigated. It was shown that such an implementation is the most effective in terms of the magnitude of the resulting errors. For this reason, let us generalize the implementation considered in \cite{Nesterova_2024} in the case of TQT.

\begin{figure}[h!]
    \centering
    \includegraphics[scale=1.0]{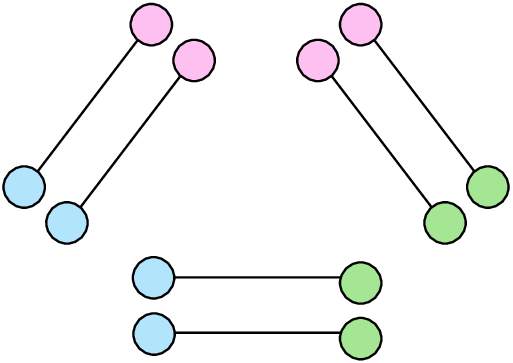}
    \caption{Six two-node cluster states distributed between Alice, Bob and Charlie. In the figure, nodes belonging to different participants are marked with different colors: pink indicates Alice’s nodes, blue indicates Bob’s nodes, and green indicates Charlie’s nodes.}
    \label{fig:six_NTCS}
\end{figure}

To realize the full functionality of TQT, we need six two-node cluster states, which are in pairs distributed between all participants. This configuration is presented in Fig. (\ref {fig:six_NTCS}).

Each two-node cluster state implements a one-directional teleportation. Therefore, this scheme is a scheme of six one-directional teleportations. As is known \cite{Nesterova_2024}, the vector of error variances of one protocol of unidirectional teleportation has the following type:
\begin{align}
    \langle \delta \hat{\vec{e}}^2_3 \rangle=\begin{pmatrix}
        2\\
        2
    \end{pmatrix} \langle \delta \hat{y}_s^2 \rangle.
\end{align}

Then the TQT in the considered implementation will have the following  vector of error variances:
\begin{align} \label{err_2}
    \langle \delta \hat{\vec{e}}^2 \rangle= \begin{pmatrix}
        2\\
        2\\
        2\\
        2\\
        2\\
        2
    \end{pmatrix}  \langle \delta \hat{y}_s^2 \rangle.
\end{align}

As we see, the error obtained in implementing several QT is always less than that obtained by using twelve nodes of the cluster state. This directly follows from the comparison of the vector of error variances (\ref{err_2}) with vectors (\ref{err_BCA}), (\ref{err_CAB}), (\ref{err_BA}), (\ref{err_cc}). 
 
We compare the vectors of error variances for several one-directional teleportations (\ref{err_2}) and three three-node cluster states (\ref{vec_3_tel}). We see that for some quadrature, an error obtained in the second case is always more. However, the error in another quadrature can be made less by choosing weight coefficients.

\section{Effectiveness of Tridirectional Teleportation Protocols in Continuous Variables}

In the work, we have presented three different ways to implement the TQT protocol in continuous variables. All three methods give the same functionality for TQT and differ only in the magnitude of errors. We evaluate the effectiveness of various methods of implementing TQT. To do this, compare their ability to correct errors in the teleportation results. This method has already been used to assess the effectiveness of the optimization of the protocols of the QT \cite{Zinatullin_2022} and the BQT \cite{Nesterova_2024} teleportation in continuous variables.

In the TQT protocol in continuous variables, we deal with small errors related to quadrature displacement. To correct these errors, an error correction protocol using GKP states was proposed in \cite{GKP}. This study demonstrated that the larger the error, the lower the probability of successfully correcting it.  In \cite{Menicucci_2014}, a method was introduced to apply these states specifically to the error correction problem in one-way quantum computing with continuous variables. Subsequently, in \cite{KOROLEV2022128149}, this error correction method was generalized to account for situations where the correction procedure itself is not ideal and may be performed with an error. Considering the results from all these studies, we can conclude that the probability of failing to correct errors in the states received by Alice, Bob, and Charlie in the TQT protocol (the probability error of the TQT protocol) is represented by the following expression:
\begin{align} \label{prot_error}
P_{err}\left(x_{er}^A,y_{er}^A,x_{er}^B,y_{er}^B,x_{er}^C,y_{er}^C\right)=1-&\mathrm{erf}\left(\frac{\sqrt{\pi}}{2\sqrt{2}\sqrt{\langle \delta \hat{y}^2_s\rangle \left(x_{er}^A+\frac{\sqrt{5}+1}{2}\right)}}\right)\mathrm{erf}\left(\frac{\sqrt{\pi}}{2\sqrt{2}\sqrt{\langle \delta \hat{y}^2_s\rangle \left(y_{er}^A+\sqrt{5} +1\right)}}\right) \nonumber\\
&\times\mathrm{erf}\left(\frac{\sqrt{\pi}}{2\sqrt{2}\sqrt{\langle \delta \hat{y}^2_s\rangle \left(x_{er}^B+\frac{\sqrt{5}+1}{2}\right)}}\right)\mathrm{erf}\left(\frac{\sqrt{\pi}}{2\sqrt{2}\sqrt{\langle \delta \hat{y}^2_s\rangle \left(y_{er}^B+\sqrt{5} +1\right)}}\right)\\
&\times\mathrm{erf}\left(\frac{\sqrt{\pi}}{2\sqrt{2}\sqrt{\langle \delta \hat{y}^2_s\rangle \left(x_{er}^C+\frac{\sqrt{5}+1}{2}\right)}}\right)\mathrm{erf}\left(\frac{\sqrt{\pi}}{2\sqrt{2}\sqrt{\langle \delta \hat{y}^2_s\rangle \left(y_{er}^C+\sqrt{5} +1\right)}}\right),
\end{align} 
where $x_{er}^{A}\langle \delta \hat{y}^2_s\rangle$, $x_{er}^{B}\langle \delta \hat{y}^2_s\rangle$ и $x_{er}^{C}\langle \delta \hat{y}^2_s\rangle$ — are the variances of the $\hat{x}$- quadrature of Alice, Bob, and Charlie,  respectively, and $y_{er}^{A}\langle \delta \hat{y}^2_s\rangle$, $y_{er}^{B}\langle \delta \hat{y}^2_s\rangle$ and$y_{er}^{C}\langle \delta \hat{y}^2_s\rangle$ — are the variances of the$\hat{y}$-quadrature of Alice, Bob and Charlie, respectively. There are two contributions to the error values in the arguments of the error function.  The first contribution is the error from the teleportation
 procedure, and the second is the error from the non-ideal error correction procedure.  The lack of symmetry in the arguments of the erf functions is due to the sequence of applying the quadratures’ correction operations. The sequence was chosen with the understanding that the compared errors (\ref{vec_3_tel}) and (\ref{err_BCA}) are larger in the $\hat{x}$-quadratures.
 
\begin{figure}[h!]
    \centering
    \includegraphics[scale=0.9]{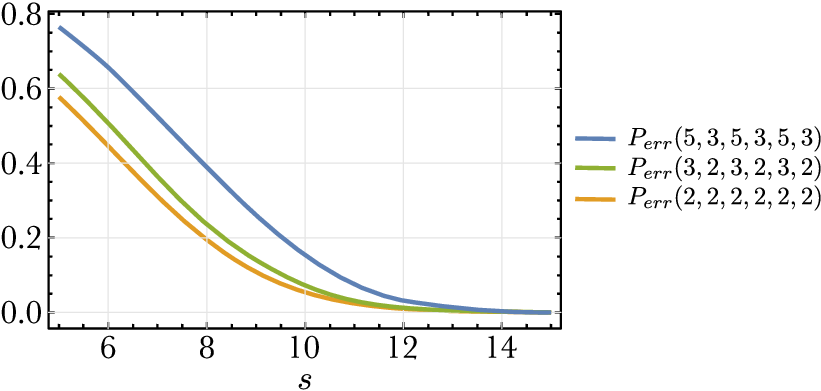}
    \caption{The probability error of the TQT protocol depending on the squeezing degree of used oscillators. Different colors indicate different cases of implementation of TQT: blue - the case of one twelve-node cluster state, orange - a case of six two-node cluster states, green - a case of three three-node cluster states. In the figure $s=-10\log_{10} 4\langle \delta \hat{y}^2_s \rangle$ is the squeezing degree of the used oscillators. The squeezing degree is measured in dB.}
    \label{fig:error}
\end{figure}

Fig. \ref{fig:error}  shows the dependence of the error probabilities of TQT protocols in different cases: with twelve-node cluster state, three three-node cluster states and six two-node cluster states. 

The weight coefficient chosen for a three-node cluster state is $g=1$. This choice is because for this weight coefficient, the function (\ref{prot_error}) for the error (\ref{vec_3_tel}) takes its smallest values. From the graph, it can be seen that the smallest probability error will be in the case where six two-node cluster states are used. The graph also shows that the importance of choosing a TQT implementation scheme increases as the squeezing degree of the used oscillators decreases. For example, when using oscillators with a squeezing degree of 8 dB, the scheme with two-node cluster states gives a 20$\%$ lower probability of incorrect error correction than a twelve-node cluster state scheme.

\section{Conclusion}
In this work, we propose three implementations of the TQT protocol in continuous variables. The key feature of our protocol is the ability to choose one of three possible TQT scenarios. In the first scenario, there is a simultaneous exchange of quantum states between all participants. The second scenario describes a pairwise exchange of states between any two teleportation participants. In the third scenario, a participant receives two quantum states from the other users.

We have shown that TQT can be implemented in continuous variables using a cluster state as the main resource. The cluster state can have different configurations (graphs). We proposed three different implementations of the TQT protocol: using a twelve-node cluster state, three three-node cluster states, and several two-node cluster states. 

Furthermore, we have shown that TQT in continuous variables is subject to errors. The variances of these errors are proportional to the variances of the squeezed quadratures of the quantum oscillators used, as well as to the weight coefficients of the cluster state graph. Since the functionality of TQT is the same for all presented configurations, the choice of the optimal implementation reduces to analyzing the errors that arise in each case. To evaluate the effectiveness, we compared the ability to correct errors in the teleportation results in each case. Based on this comparison, we found that the use of set two-node cluster states results in the smallest probability error, making it the most suitable for the TQT protocol. This result is easily explained from a physical point of view. During teleportation, each squeezed quantum oscillator acts as a source of error, so minimizing the error requires using the smallest number of oscillators. Although the total number of squeezed oscillators in this case is larger than in TQT with three three-node cluster states, the number of squeezed oscillators used during the protocol is smaller. In addition, the two-node cluster state protocol is simpler from an experimental point of view, as it does not require the creation of complex entanglement states.

\vspace{0.5 cm}

This research was supported by the Theoretical Physics and Mathematics Advancement Foundation "BASIS" (Grants No. 24-1-3-14-1). E.A. Nеsterova acknowledges Mikhail Romanovsky for his financial support. 

\bibliography{bibliography}

\end{document}